\documentclass{aa}
\usepackage{natbib,epsfig,txfonts,graphicx}

\def \he  {HeI~$\lambda$5876\,}
\def \hee {HeI~$\lambda$6678\,}

\def\Teff{$T_{\rm eff}$}
\def\logg{$\log g$}

\def\Msun {${\rm M}_{\odot}$}

\def\Rstar{$R_{\ast}$}

\def\Mdot{${\dot M}$}

\def\vesini{$v_e\,$sin($i$)}

\def\vinf {$v_{\rm \infty}$}

\def\vsys{$v_{\rm sys}$\,}

\def\Ha {H$_{\rm \alpha}$}

\def\kms {km~s$^{-1}$}

\def\beq{\begin{equation}}
\def\eeq{\end{equation}}
\def\beqa{\begin{eqnarray}}
\def\eeqa{\end{eqnarray}}

\begin{document}

\title{ The superimposed photospheric and stellar wind
variability of the O-type supergiant $\alpha$ Cam}


\author{R.K.Prinja\inst{1}, N.Markova\inst{2}, S. Scuderi\inst{3},  
and H. Markov\inst{2}}

\offprints{R.K. Prinja}

\institute{Department of Physics {\&} Astronomy, University
College London, Gower Street, London WC1E 6BT, U.K.\email{rkp@star.ucl.ac.uk}
\and
Institute of Astronomy, Bulgarian National Astronomical Observatory,
  P.O. Box 136, 4700 Smoljan, Bulgaria, \email{nmarkova@astro.bas.bg, hmarkov@astro.bas.bg}
\and
INAF - Osservatorio Astrofisico di Catania, Via S. Sofia 78, I-95123 Catania, 
Italy, \email{scuderi@oact.inaf.it}
}

\date{Received; accepted }

\abstract{}
{
This study {{seeks to provide empirical constraints on the}} different 
physical
components that can {{act to yield}} temporal variability in 
predominantly or 
partially wind-formed optical lines of luminous OB stars, and thus potentially 
affect the reliable determination of fundamental parameters, {{
including mass-loss rates via clumped winds.}}
}
{Using time-series spectroscopy from epochs spread over $\sim$ 4 years,
we present a case study of the O9.5 supergiant $\alpha$ Cam. We demonstrate 
that the \he (2$^3$P$^0$--3$^3$D) line is an important diagnostic for photospheric 
and wind variability in this star. The actions of large radial velocity shifts 
(up to $\sim$ 30 km s$^{-1}$) in the photospheric absorption lines can also 
affect the morphology of the \Ha line profile, which is commonly used for
measuring mass-loss rates in massive stars.}
{We identify a 0.36-day period in subtle absorption profile changes in \he,
which likely betrays {{photospheric structure, perhaps due
to low-order non-radial pulsations.}} 
This signal persist over $\sim$ 2 months, but it is not present
2 years later (November 2004); it is also not seen in the stellar wind
components of the line profiles. Using a pure \Ha line-synthesis
code we interpret maximum changes in the red-ward and peak emission
of $\alpha$ Cam in terms of mass-loss rate differences in the
range $\sim$ 5.1 $\times$ 10$^{-6}$ to 6.5 $\times$ 10$^{-6}$
M$_\odot$ yr$^{-1}$. However, the models generally fail to reproduce
the morphology of blueward (possibly absorptive) regions of the profiles.}
{The optical line profiles of $\alpha$ Cam are affected by (i)
{{deep-seated fluctuations close to, or at, the photosphere}}, (ii) 
atmospheric velocity
gradients, and (iii) large-scale stellar wind structure.
This study provides new empirical perspectives on accurate line-synthesis
modelling of stellar wind signatures in massive luminous stars.}


\titlerunning{The variable signatures of $\alpha$ Cam}
\authorrunning{Prinja et al.}

\maketitle
   \maketitle
%

\section{Introduction}

Through their powerful stellar winds, high ionizing fluxes, and ultimate 
demise as supernovae, massive stars have a profound effect on the 
dynamical and chemical evolution of our own and other galaxies. The hot 
star winds are driven by line scattering of radiation, and the extent of the 
mass-loss is a major factor in the evolution of these stars; it can for
 instance influence whether the end-state is a neutron star or a black 
hole \citep[e.g.][]{Heger03}.  Therefore it is important that the nature 
of 
these outflows is well understood, and in particular the mass-loss rates 
are reliably determined.

One of the most serious obstacles in this path arises from detailed 
optical and ultraviolet observations that reveal widespread line profile
variability, implying clumped structure in the winds
\citep[see][plus references within]{Prinja02, Markova04, Morel04}.
The developing picture of massive star winds as a clumped
medium is further supported by recent model atmosphere calculations;
e.g. \citet{Bouret05} need to include substantial clumping
factors in their models to match key wind features in the spectra
of O-type stars, implying significant downward revisions in the 
canonical mass-loss rates.
Clumping may be a consequence of the inherent instability of the
radiation driving mechanism of the wind, or of the creation of
coherent structures arising from photospheric disturbances, or both.
In either case, wind clumping is inextricably linked to the
fundamental physical problem of radiation driven winds.

Our objective in the study presented here is to exploit detailed
time-series optical observations of the late-O supergiant
$\alpha$ Cam (HD~30614) to provide empirical constraints for
{\it ab initio} model atmosphere codes that are commonly used to
synthesise spectra in order to derive fundamental stellar and wind
parameters (e.g \citealt[CMFGEN] {Hillier1998},\citealt[WMbasic] 
{Pauldrach01}, 
\citealt[FASTWIND]{Puls05}). Our 
approach is to
constrain the components that contribute to changes evident in
diagnostic wind and photospheric lines. We seek to `decompose' the
variability patterns and assess in particular the degree to which
fluctuations in the wind-formed features are due to `contamination' by
photospheric and deep-seated atmospheric activity. This is
particularly critical in late O and early B stars, where the
\Ha \,emission may be weak; see e.g. the imprints of
photospheric signals seen in the Balmer lines of $\epsilon$ Ori
\citep[B0 Ia,]{Prinja04}. Our study of $\alpha$ Cam is also
designed to probe causal links between photospheric disturbances
and spatial wind structure.

The target $\alpha$ Cam is previously known to exhibit 
line-profile variability in \Ha \, (e.g. \citet{Ebbets82, Fullerton1990, 
Kaper1997}) as well as in some other absorption lines in 
the optical spectrum \citep{Fullerton96, Markova02}.
Whist its UV resonance lines are strong and saturated,
\citet{Lamers1988} reported on weak low-velocity fluctuations and 
changes in
the soft blue wings of saturated (`black') absorption troughs.
Some fundamental parameters of $\alpha$ Cam are listed in Table~\ref{para}.

\begin{table}
\centering
\caption{Fundamental parameters for $\alpha$ Cam.}
\label{para}
\begin{tabular}{l@{\hspace{2mm}}l@{\hspace{2mm}}l@{\hspace{2mm}}}
   Parameter & Value & Reference\\
\hline
\hline
\\
Spectral Type & O9.5 Ia & \citet{Walborn1973} \\
$T_{\rm eff}$ & 29000 & \citet{Repolust04} \\
$R_{\star}/R_\odot$ & 32.5 & \citet{Repolust04} \\
\vesini & 115 km s$^{-1}$ & \citet{Penny1996} \\
$M_{\star}/M_\odot$ & 20 & \citet{Schaller1992} \\
$V_\infty$ & 1550 km s$^{-1}$ & \citet{Haser1995} \\
$P_{\rm rot}(max)$ & $\sim$ 14.3 days & \\
\hline
\end{tabular}
\end{table}

   \begin{figure*}
   \centering
      \includegraphics[scale=0.67]{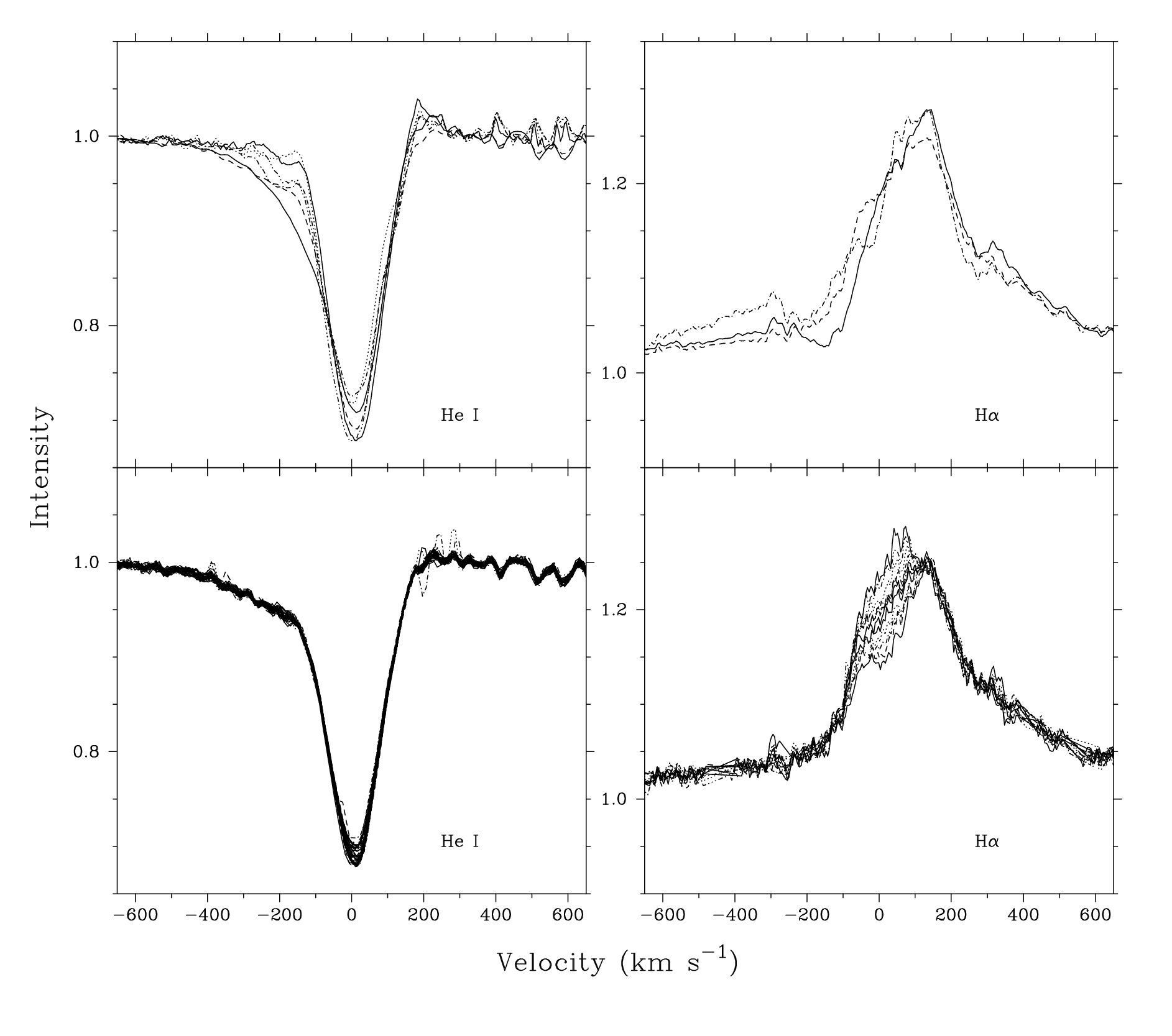}
   \caption{Examples of variability in the \he and \Ha line 
   profiles of $\alpha$ Cam are shown for two characteristic time-scales; 
   (i) night-to-night mean profiles (upper panels) and 
   (ii) $\sim$ hourly changes within a single night (lower panels).}
              \label{fig1}%
    \end{figure*}
%

\section{Observations}

Our investigations are based on three datasets; (i) time-series
of primarily He~I $\lambda$5875.67 obtained during January, February
and March 2002, (ii) time-series of \Ha, \hee and \he obtained 
during November 2004, and (iii) comparisons of (i) and (ii) to \Ha 
observations secured in 1999 and described by \citet{Markova02}. A 
summary 
log of the observations of $\alpha$ Cam is given in Table~\ref{log}. 

The majority of the 2002 data were obtained using the Coud\'e spectrograph 
of the 2-m telescope at the National Astronomical Observatory, NAO (Bulgaria) 
equipped with a PHOTOMETRICS CCD (1024$\times$1024, 24$\mu$). 
\footnote{This detector was characterized by an $rms$ read-out noise 
3.3 electrons per pixel (2.7 ADU with 1.21 electrons per ADU)}
The spectrograph was used in two different configurations producing
spectra with a reciprocal dispersion of $\sim$0.2 \AA\ pixel$^{-1}$
(effective resolution R=15\,000) and of $\sim$0.1 \AA\ pixel$^{-1}$ 
(effective resolution R=30\,000) over a wavelength range of respectively, 
$\sim$200 and $\sim$ 100 \AA\,.  The typical signal-to-noise of the
individual spectra, adjacent to \he, ranges  between 100 to 500.

The 2002 time-series from NAO is complemented by observations obtained 
between 21 to 23 January, 2002 from the 0.9-m telescope at Catania 
Observatory (CO, Sicily).  The setup employed 
 an Echelle spectrograph and a SITe CCD detector, to yield a spectral 
resolution R $\sim$ 20\,000 and a typical signal-to-noise of $\sim$ 130.

The 2004 time-series (Table 2) were also secured at the NAO 2-m
telescope employing the Coud\'e spectrograph to cover 
wavelength regions including \he, \Ha\, and \hee. 
The effective spectral  resolution of these data is 15\,000 with 
signal-to-noise  (adjacent to \he) between 300 to 400.

The data from both observatories were reduced in a homogeneous 
manner, with standard procedures for bias subtraction, flat-fielding 
and wavelength calibration.  The line profiles of $\alpha$ Cam were 
subsequently normalised by fitting a low-order polynomial through selected 
continuum 
windows. Finally, the telluric water vapour lines were removed by dividing 
individual spectra with a scaled model telluric spectrum
\citep[e.g.][]{Markova00}. Note that the internal stability of the 
velocity scale in the extracted spectra  (specified by the 
standard deviation in velocity of the IS NaI D line at 
$\lambda$ 5889.95)  is less then 1 \kms for both the NAO and 
CO data sets. Throughout the paper 
we have corrected to the stellar rest frame for a radial velocity 
of 6 km s$^{-1}$ \citep[e.g.][]{HJ1982}. 
 
The intensive \he time-series of 114 spectra obtained between 21 to 
27 January 2002 is therefore our primary dataset for probing short-term 
modulated or periodic line profile variability.


\begin{table}
 \centering
  \caption{Summary log of observation.}
 
\begin{tabular}{l@{\hspace{2mm}}l@{\hspace{2mm}}l@{\hspace{2mm}}c@{\hspace{2mm}}}
   Observatory & UT Date          & HJD 2~450~000 +     & No. of \\
               &                  &                     & spectra \\
\\
Catania 0.9m & 2002 Jan. 21,22,23 & 2296.400$-$2298.577 & 50 \\
NAO 2m       & 2002 Jan. 25,26,27 & 2300.240$-$2301.396 & 64 \\
NAO 2m       & 2002 Feb. 25,26    & 2331.246$-$2332.486 & 36 \\
NAO 2m       & 2002 March 2,3     & 2336.220$-$2337.433 & 30 \\
NAO 2m       & 2004 Nov. 26,27,28 & 3336.367$-$3338.654 & 29 \\
\end{tabular}
\label{log}
\end{table}

\section{Variability characteristics}

With the possible exception of \Ha, the optical spectrum of $\alpha$~Cam 
is typical for its
late O spectral class, with (mostly) symmetrical He~I absorption
lines, very weak metal lines (e.g. C~IV $\lambda\lambda$5801, 5811),
and weak He~II $\lambda$4686 (absorption). We are confident that
ours is a case study of a `normal' late O-type supergiant and focus in
this section on the temporal behaviour of \he and \Ha.

   \begin{figure}
   \centering
      \includegraphics[scale=0.47]{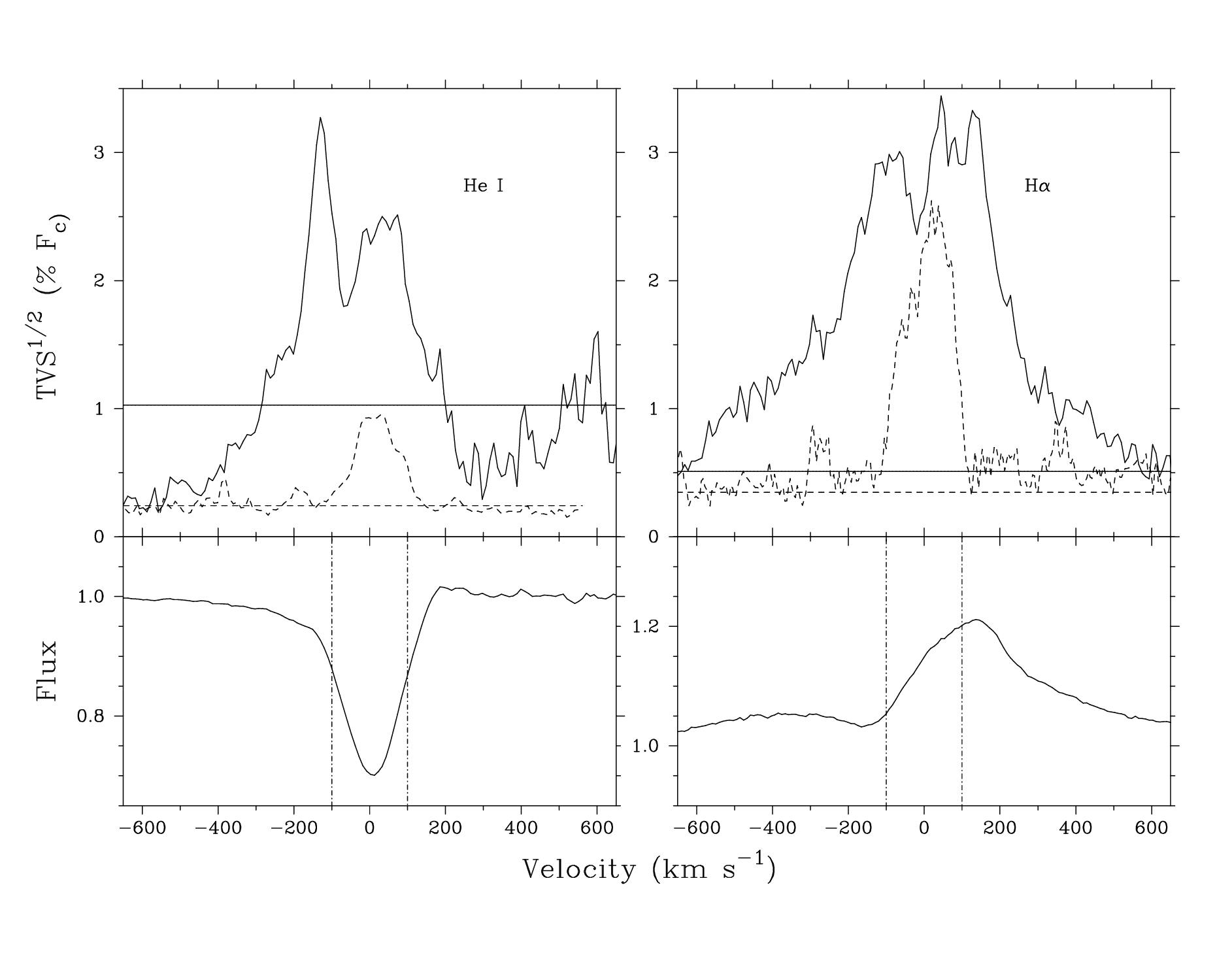}
   \caption{The temporal variance spectrum (TVS) for the cases 
     of night-to-night 
     changes (solid lines)
     and $\sim$ hourly 
     changes (dashed lines).
{The horizontal lines in the upper panels indicate the
respective 95{\%} confidence levels. The \he fluctuations
reward of +500 km s$^{-1}$ are an artifact from the
removal of telluric lines.}
     The vertical lines (bottom panel) mark $\pm$\vesini.}
              \label{fig2}%
    \end{figure}
%

\subsection{\he (2$^3$P$^0$--3$^3$D)}

This is an important diagnostic line, since the lower level of the 
transition can become substantially populated in the spectra of 
massive stars, which then effectively becomes the ground state and 
the line may reveal P~Cygni-like signatures. 

The night-to-night variations are contrasted against shorter time-scale 
(hourly) changes in Fig.~1. Even though \he is predominately a photospheric 
line in $\alpha$~Cam,  the night-to-night mean profiles in Fig.~1 
(upper-left panel) reveal evidence of a variable stellar wind component, 
with absorptive changes in the blue-wing which contrast with the less 
extended, mostly `stable' red-wing and a weak emission. The data indicate 
that the deep-seated inner wind region ($\le$ 2 R$_{\star}$) of 
the star is not steady 
This night-to-night variability is accompanied by more rapid and subtle 
changes (down to $\sim$ 2-3{\%} of the continuum level)  in the core 
absorption. This behaviour is seen in intensive (hourly) time-series 
captured during individual nights (Fig.~1; lower-left panel)

To quantify the significance level of the line profile variability
as a function of velocity, we have determined the Temporal Variance
Spectrum (TVS) according to the methods of \citet{Fullerton96}.
The results are shown in Fig. 2 (left-hand panel). The night-to-night
stellar wind variation in \he extends over $\sim$ $-$500 to +200 \kms 
(i.e. $\sim-$0.3 $v_\infty$ to +0.13 $v_\infty$). The $\sim$ hourly 
core absorption changes are essentially symmetric about rest velocity 
and constrained within the boundary marked by $\pm$ \vesini 
thus suggesting the presence of photospheric activity in the star.

\subsection{\Ha}

The corresponding results for \Ha are shown in the right-hand
panels of Figs. 1 and 2.  The mean \Ha profile (e.g. Fig. 2) consists 
of a well-developed P Cygni-like core superimposed on broad 
emission wings.
In O-type supergiants of same spectral type, \Ha\ Profiles 
with similar morphology to $\alpha$~Cam 
are indicative of relatively high wind density connected to a higher 
mass-loss rate (e.g. \citealt{Markova05}).
The well developed \Ha\, profile observed here may 
(in part) reflect the Ia luminosity class, and perhaps also the 
presence of clumps in the outflow.

From Figs~1 and 2 it is also obvious that the emission peak of \Ha\, in
$\alpha$~Cam is red-shifted, typically occurring at +100 km s$^{-1}$.
Red-shifted emission has also been observed in UV resonance
lines of O-type stars, as well as in \Ha. These two phenomena, 
though
similar on a first glance, seem to have different interpretations. While
in the former case the red-shifted emission can be explained in terms of
``micro-turbulence effects'' with $v_{micro}$\, of the order of 0.01\vinf
\citep{Hamann80}, in the later case it likely results from the
interaction between the red-wing of the Stark-broadened photospheric
profile and wind emission of certain strength, namely $log <\rho>$ $>$
13.6 \footnote{ $<\rho>$ is the ``mean wind density''calculated at a
typical location of 1.4~\Rstar} \citep{Markova05}.  Indeed, with its
``mean wind density'', $log <\rho>$, =13.64, $\alpha$~Cam is just above 
this
limit and thus consistent with the interpretation suggested by Markova et
al.

The TVS for \Ha\, $nightly$ means (Fig. 2) extends almost
symmetrically in velocity space (blue and red limits at about 
$\pm$600 \kms, i.e. 0.39\vinf) with blueward amplitudes (with respect to 
the rest wavelength) slightly stronger then the red-ward  ones.
As can be seen from a comparison to Figure 1, the observed blue-to-red 
asymmetry in  the strength of the TVS amplitudes is more likely due 
to {\it absorptive}  blue-ward \Ha features, perhaps caused by wind 
structure localised along the line-of-sight to the stellar disk. In contrast, 
the overall emission profile is produced by gas occupying a greater 
volume than the gas responsible for (episodic) absorption features.

Generally the \Ha\, profile of $\alpha$~Cam is not substantially
variable on $hourly$ time-scales. However, an exceptional case is
shown in Fig. 1 (lower-right panel), where localised systematic
changes are clearly evident over $\sim$ 10 hours. These intriguing 
variations are tightly constrained within $\pm v_e\,$sin($i$) (see Fig. 2). 
Such a behaviour has previously been documented by \citet{Markova02} and
we discuss it further in Sect. 3.3.2.

In a given epoch, over nightly time-scales the total \Ha\, equivalent width 
may vary by up to 10{\%}; however, differences of 30$-$40 \% are also 
noted between line profiles secured in different years (see Sect. 4).

   \begin{figure}
   \centering
      \includegraphics[scale=0.67]{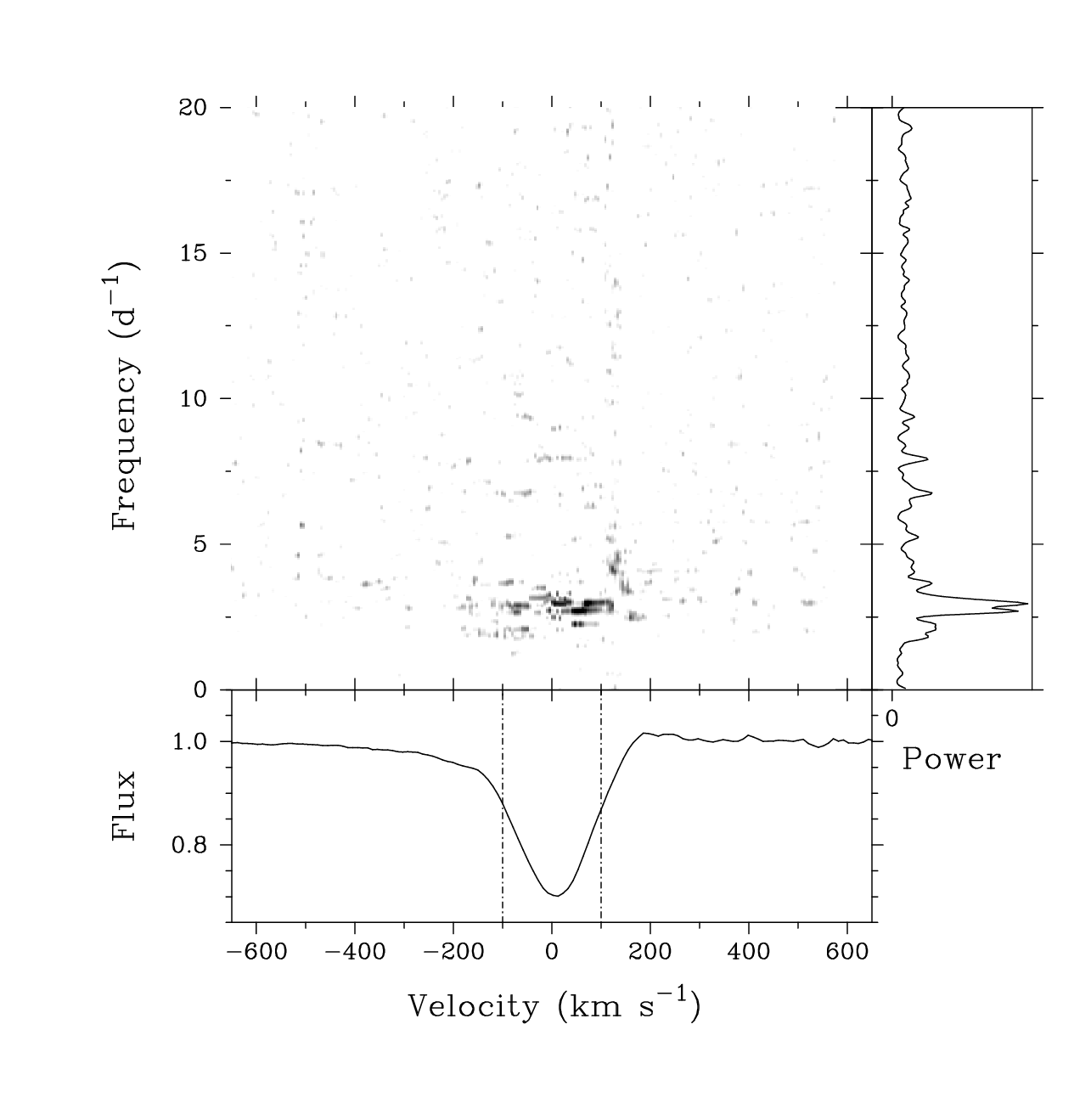}
   \caption{Two-dimensional representation of Fourier analysis
      of the \he line profile variability. The significant period in 
      the power spectrum (right-hand panel) is at $\sim$ 0.36-days. 
      (The vertical lines mark $\pm v_e\,$sin($i$).)}
              \label{fig3}%
    \end{figure}
%

   \begin{figure*}
   \centering
      \includegraphics[scale=0.73]{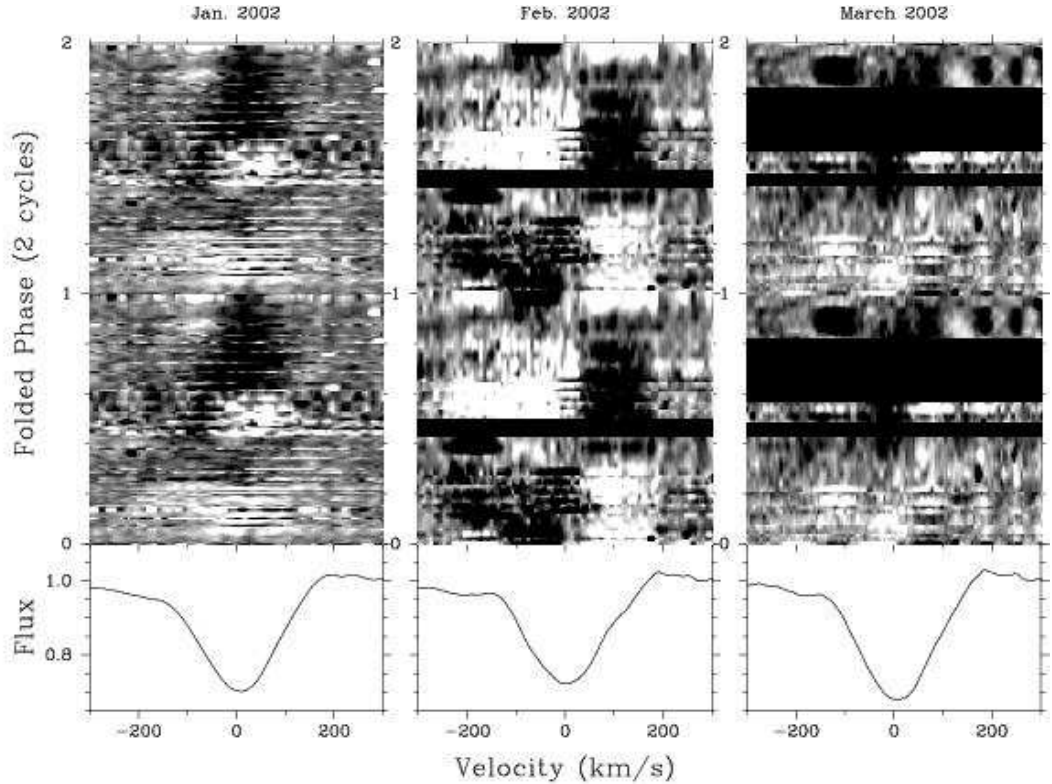}
   \caption{Grey-scale representations of variability in
    \he for individual spectra phased on the 0.36-d period.
{Individual quotient spectra with respect to the mean are
shown, such that darker shades indicate regions of enhanced
absorption (reduced flux) compared to the mean. The total range
of intensities plotted is 0.99 (black) to 1.01 (white).}
}
              \label{fig4}%
    \end{figure*}
%

\subsection{Photospheric activity}

\subsubsection{The 2002 time series}

It is clear from Figs. 1 and 2 that the temporal stellar wind
activity in $\alpha$~Cam dominates the overall variance levels
of \he and \Ha. Since the wind changes occur over time-scales 
of days, we are still able to pursue the effects of shorter 
time-scale {\it photospheric} variability in these lines (particularly 
in \he). To emphasis the photospheric changes, the individual line 
profiles were normalised by the corresponding nightly mean profile,
{to `remove' the larger amplitude night-to-night changes}, 
and time-series analyses was then carried out on the quotient spectra.

The most suitable dataset for a search for short-period systematic
variability is the intensive \he time-series obtained from Catania and 
the Bulgarian NAO between 21 to 27 January, 2002, totalling 114 spectra 
(Table 2; note that \Ha was not observed during this setup). We 
carried out a periodogram analysis of these data, using the CLEAN algorithm 
\citep{Roberts1987} to deconvolve the features of the window function 
from the discrete Fourier transform. The two-dimensional grey-scale 
representation of the periodogram for He~I is shown in Fig. 3 (for a 
gain=0.5 and 100 iterations;
{lower values of gain or greater iterations yield
essentially the same power spectrum}). Though the power 
across the absorption 
core is slightly dis-jointed, the {results indicate a (dominant) 
primary frequency of $\sim$ 2.78 days$^{-1}$, corresponding to a period of 
0.36 days.}
(Note this period
is consistent with the modulations reported by \citet{Markova02} in
data for \hee secured in late 1998 and early 1999.)
We are very confident that 0.36-d period is not directly connected to
the sampling window of each night, where the length of the nightly
runs in 2002 is between $\sim$ 0.12 to 0.27 days (Table 2).
Note, that a Fourier analysis of the original He~I data (i.e. not
residual with respect to the nightly mean) {reveals essentially the
same modulation signal as in Fig. 3, but does not show} 
any
significant power on this (or any other) period in the
{more blue-ward} region
dominated by night-to-night {\it wind} variability, i.e. over $\sim$ $-$100 
to $-$500 km s$^{-1}$.

Grey-scale images of the phase versus velocity behaviour of
the individual residual He~I spectra on the 0.36-d period are
shown in Fig. 4. The left-hand panel show the coherent behaviour
during our primary (January 2002) time-series. The sparser  datasets 
secured in February 2002 and March 2002 (Table 2) are not suitable 
for a reliable Fourier analysis, but
{they do provide some indication that systematic modulations
persist for several weeks over comparable timescales, though not
with the same phase relationship.}

The data from January 2002 reveal some indication for progressive
changes across the absorption profile, though the acceleration
of the pseudo-absorption (emission) feature is not substantial.
{There is evidence for characteristic `blue-to-red' motion 
(see Fig. 5)}, accommodated within the projected 
rotation
velocity (115 km s$^{-1}$). We estimate a prograde 
feature travelling across the line centre with (dV/d$\phi$)
$\sim$ 80 km s$^{-1}$/cycle ({Fig. 5}). This behaviour
{would be consistent with a low-order sectorial} non-radial pulsation
mode. These time-series of $\alpha$ Cam are however not extensive
enough, nor sufficiently high signal-to-noise, to attempt more
detailed modelling to determine pulsational parameters.

   \begin{figure}
   \centering
      \includegraphics[scale=0.47]{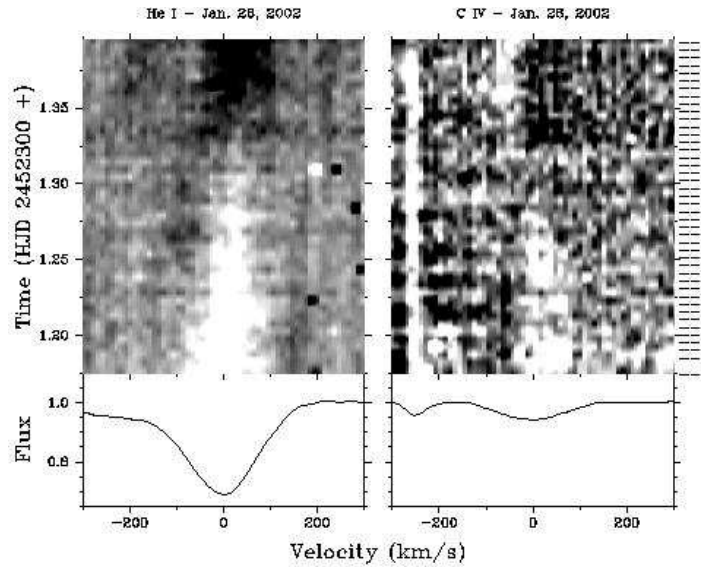}
   \caption{An example of absorption line profile changes in
\he and C~IV $\lambda$5801 over $\sim$ 5.3 hours. 
{Individual
quotients with respect the mean are plotted, for the same
dynamical scale as in Fig. 4. The tick-marks along the right-hand
side indicate the timings of the individual spectra, for both
panels.}
}
              \label{fig5}%
    \end{figure}
%

The wavelength range of the Bulgarian NAO data of January 2002
includes the very weak photospheric metal lines of
C~IV $\lambda\lambda$5801, 5812. Unfortunately, the intensity of these 
lines is rather small but we are confident that 
they are also temporally active. The C~IV $\lambda$5801 data for 
26 January, 2002 covering $\sim$ 5.3 hours is shown as a grey-scale 
(`dynamic') representation in Fig. 5, together with the corresponding \he 
line profiles. The subtle prograde travelling pattern identified above in
He~I is {tentatively mimicked} in C~IV, which
supports an interpretation in terms of photospheric velocity fields;
{though confirmation of this result clearly requires
much higher signal-to-noise time-series data.}

\subsubsection{The 2004 time series}

Interestingly, the 0.36-d period derived from the 2002 data of
$\alpha$~Cam is {\it not} present in the observations secured
during November 2004. The line profile changes are generally
weaker in 2004 and the Fourier analysis does not reveal any
significant frequencies in \he, \Ha\, or \hee.
However, whilst the time-series of 26 and 28 November, 2004 are
almost steady, a different episodic property is evident
during 27 November, 2004. In contrast to the weak fluctuations
within the absorption trough seen in 2002 (Figs. 4 and 5), a
systematic velocity shift occurs in the absorption components
of the He~I lines during this night; this behaviour is shown
in Fig.~6.

   \begin{figure*}
   \centering
      \includegraphics[scale=0.73]{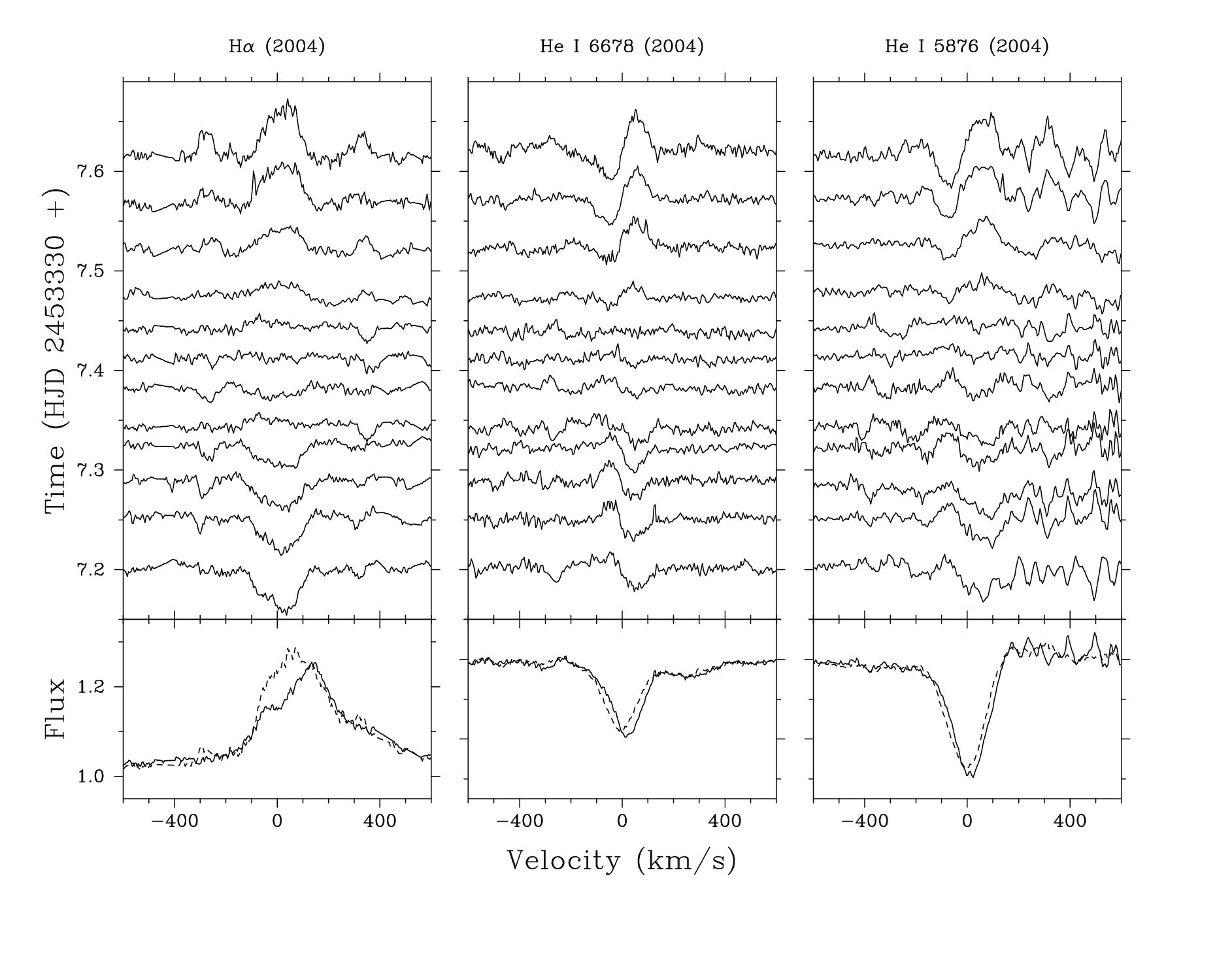}
   \caption{
{Upper panels: Montage of quotient spectra with respect
to the mean, showing variability} in \Ha,
\hee and \he during 27 November 2004.
{Lower panels: Representative examples of the original (not
residual) line profiles.}
Note the radial velocity shifts in the He lines and the
corresponding change in \Ha peak emission flux.
{(For \he the Na I D interstellar lines (Sect. 2) provide
an excellent wavelength fiducial.)}
}
              \label{fig6}%
    \end{figure*}
%

The central velocity (and wings) of \hee shifts by up to $\sim$ 30 
km s$^{-1}$ during the night, while the motion of \he has a peak shift
of $\sim$ 15 km s$^{-1}$ (but in a corresponding manner). Interestingly, 
these changes in absorption velocity are accompanied by relatively 
strong changes in the \Ha\, emission flux, which are centred at 
rest velocity and occur within $v_e\,$sin($i$) (see Fig. 6).
The \Ha\, flux is a maximum in this region when the
He~I absorption lines have their greatest blue-ward motion.

We stress that the behaviour documented in Fig. 6 is isolated to  
November 27, 2004,  it was not witnessed for example during 2002 or 
during the remaining nights in 2004.  This behaviour (also shown here 
in Fig. 1) is empirically very  different from the broader changes due 
to stellar wind variability that grossly affect the blue and red wings 
of \Ha\, and might in principle be due either to variations in 
radial velocity of the underlying photospheric profiles or to localised 
changes in density stratification in the innermost part of the wind.

To test  the first possibility we employed the 
approximate method developed by \citet{Puls1996} \citep[see also][and 
applications in Sect. 4]{Markova04} to calculate a pair of `representative' 
 \Ha\, model profiles: One with a normal TLUSTY (\Teff = 30\,000K, \logg = 3.2)
photospheric profile and the other with the same TLUSTY input
but now shifted in radial velocity by $-$30 km s$^{-1}$
(cf. \hee in Fig.~6). The corresponding model \Ha\,
profiles are shown in Fig.~7, demonstrating some correspondence
to the observed \Ha\, changes shown in Fig. 6. The effect
of shifting the core \Ha\, photospheric absorption blue-wards
is (naturally) to reveal more of the central \Ha\, emission.

It seems therefore possible that  the photospheric lines of 
$\alpha$ ~Cam are affected by large-amplitude radial velocity shifts. It is unknown 
whether this characteristic is periodic over a much longer time-scale, 
but the phenomenon was only seen once in our 2002 and 2004 data sets. 
The radial velocity motion is differential, such that the extent of the 
velocity shifts is different for lines formed at varying depths 
in the atmosphere. This likely reflects a velocity gradient in the atmosphere, 
though the physical origin of this deep-seated disturbance is not known.

On the other hand,   the \Ha\, variability pattern illustrated  in the lower 
right  panel of Fig.~1 seems to be quite similar to the ones observed  
by Markova  on Dec. 31, 1998 and on Jan 6, 1999 \citep[see Fig.4 ]{Markova02}.
In addition,  the radial velocity behaviour of  the \hee line during those two 
nights in 1998 and 1999 was similar to that shown in Fig 6, i.e.  the increase 
in the \Ha\, emission flux localised within $\pm$ \vesini \,was accompanied 
by a shift to the blue in the radial velocity of the line. The time 
resolution of the 
1998-1999 data set was  very low (3 spectra per night  only) and thus 
prevents a more detailed comparison of the phenomena to the 2004 event.
Nevertheless, the empirical similarity of the Nov. 27, 2004 
variations and those observed on Dec. 31, 1989 supports a common
physical origin, perhaps in terms of large-scale perturbations in
the inner-most wind regions, as suggested by \citet{Markova02}.

\section{Stellar wind characteristics}

It is well known that the stellar winds of OB stars are highly variable. Evidence 
for wind structure is seen in modulated UV and optical lines (e.g. 
\citealt{Kaper1996, deJong01,Prinja04, Markova05}), excess flux at
infrared and millimetre wavelengths \citep{Runacres1996}, and detection 
of X-rays \citep[e.g.][]{Chlebowski1989, Cassinelli01}. $\alpha$~Cam is no 
exception in this respect as already demonstrated in Figs. 1 and 2 (see also
references in Sect. 1).

The \he line of $\alpha$ Cam offers a weak diagnostic of variable 
wind conditions via fluctuations in the blue-ward absorption wing 
(Fig. 1); unfortunately we only have corresponding \Ha data
 for the November 2004 run. Excluding the exceptional behaviour 
seen on 27 November, 2004 (Sect. 3.1), a maximum increase in 
\Ha\, equivalent width between $-$100 to $-$600 \kms of 
$\sim$ 0.4{\AA} is accompanied by a 0.05{\AA} {\it absorption} 
enhancement in He~I. We do not find any evidence for periodic or 
modulated stellar wind changes over $\sim$ hourly time-scales. 
Our data set is not suitable for an assessment of temporal behaviour 
on stellar rotational time-scales.

We focus in this section instead on the large-scale, epoch-to-epoch
changes evident in \Ha, which are, of course, superimposed
on the more subtle photospheric and deep atmospheric structure
described in Sect. 3. Our goal here is to explore extreme \Ha\, 
morphologies in $\alpha$~Cam that provide challenges for line-synthesis 
modelling and likely reflect variable density distributions in the 
inner-wind regions. 

The overall \Ha line profile shape can vary substantially over 
yearly time-scales. A particular example is shown in Fig. 8 (upper panel) 
where we compare line profiles from November 2004 and January 1999. 
The $\sim$ 30{\%} change in equivalent width is mostly around line centre 
and toward the red-ward emission wing. To fit the two profiles we employed 
again the approximate method of  \citet{Puls1996}, modified by 
\citet{Markova04}, to account for the effects of line-blocking 
and blanketing.  We find 
that the observed red-wing and peak are reasonably well matched for fixed 
terminal velocity = 1550 km s$^{-1}$, $\beta$ velocity law index = 1.05 and 
varying the mass-loss rate (\Mdot) between 
6.5 $\times$ 10$^{-6}$ \Msun yr$^{-1}$to 5.1 $\times$ 10$^{-6}$\Msun yr$^{-1}$. 
The model with the lowered emission flux cannot however reproduce the 
blue-ward regions of the line profile from 1999, since it under-predicts the 
emission in this region.  

The lower panel in Fig. 8 shows a markedly different temporal
characteristic in the \Ha\, profile of $\alpha$~Cam. These two spectra
were taken 4 days apart in January 1999, and they exhibit broadly similar
red-ward emission and peak fluxes. However, in this case the
{\it blue-ward}
regions of the profile are highly variable between $\sim$ $-$50 to $-$450
km s$^{-1}$ (i.e. well beyond the domain of He~II $\lambda$6560.2). Whilst
models with \Mdot\, differing between $\sim$6.2 $\times$ 10$^{-6}$ \Msun 
yr$^{-1}$ and
5.9 $\times$ 10$^{-6}$ \Msun yr$^{-1}$ ($v_\infty$ = 1550, $\beta$ = 1.05)
can broadly match the red-wing, they once again cannot reproduce the
differing morphologies of the blue-ward regions.

We note therefore that though the variable red-ward
\Ha wings allow some constraint on \Mdot\, and $\beta$, the blue-ward
fluctuations pose a more serious discrepancy for spherically symmetric
stellar wind models with smooth density stratification.

\section{Discussion}

{We have demonstrated in this study that the primarily, or
partially, wind-formed spectral lines of the late O supergiant
$\alpha$ Cam can exhibit variability that arises in different
physical regimes of the star. This `contamination' of wind signatures
by variable non-wind components is directly relevant to line-synthesis
modelling, particularly in the context of revising mass-loss rates
due to the presence of small- and large-scale structure in the
outflows.}
In particular,
we have found that the temporal behaviour of $\alpha$ Cam's optical 
spectrum is 
characterised by at least three different variability patters:
\begin{enumerate}
\item[i)] {systematic changes in the photospheric lines, with
some evidence that the behaviour may be linked to surface velocity
fields due to non-radial pulsations},
\item[ii)] short time-scale ($\sim$ hourly) perturbations that
differentially shift the radial velocity of He~I 
lines, and potentially give rise to localised changes in \Ha emission, 
\item[iii)] a variable stellar wind, which may be parameterised
typically in terms of a $\sim$ 5{\%} change in mass-loss rate over several 
days, but very occasionally this difference can rise to $\sim$ 30{\%}.
The red-ward emission changes are accompanied by more enigmatic
blue-ward changes that may result from absorption
effects linked to localised density structures in the outflow.
\end{enumerate}

   \begin{figure}
   \centering
      \includegraphics[scale=0.67]{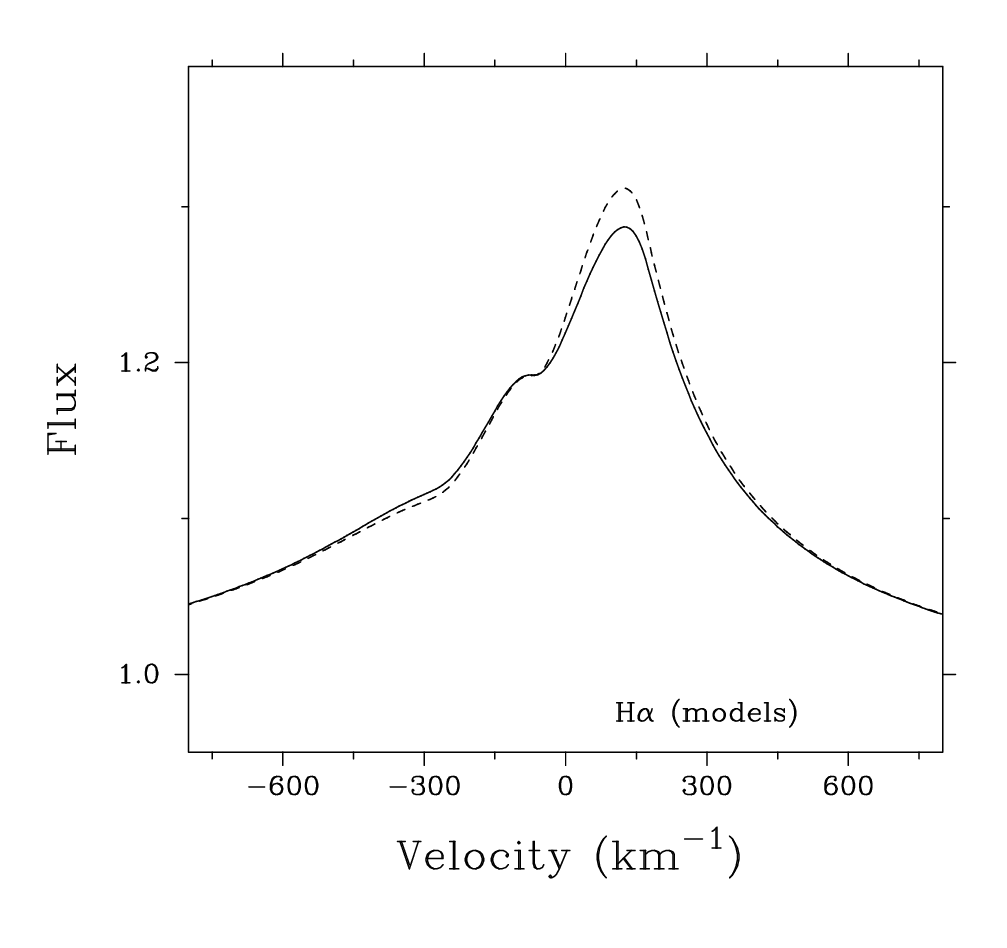}
   \caption{`Representative' model \Ha profiles, demonstrating
the effect of shifting the TLUSTY photospheric models by
$-$30 km s$^{-1}$ (dashed line; see Sect. 3.3).}
              \label{fig7}%
    \end{figure}
%

   \begin{figure}
   \centering
      \includegraphics[scale=0.67]{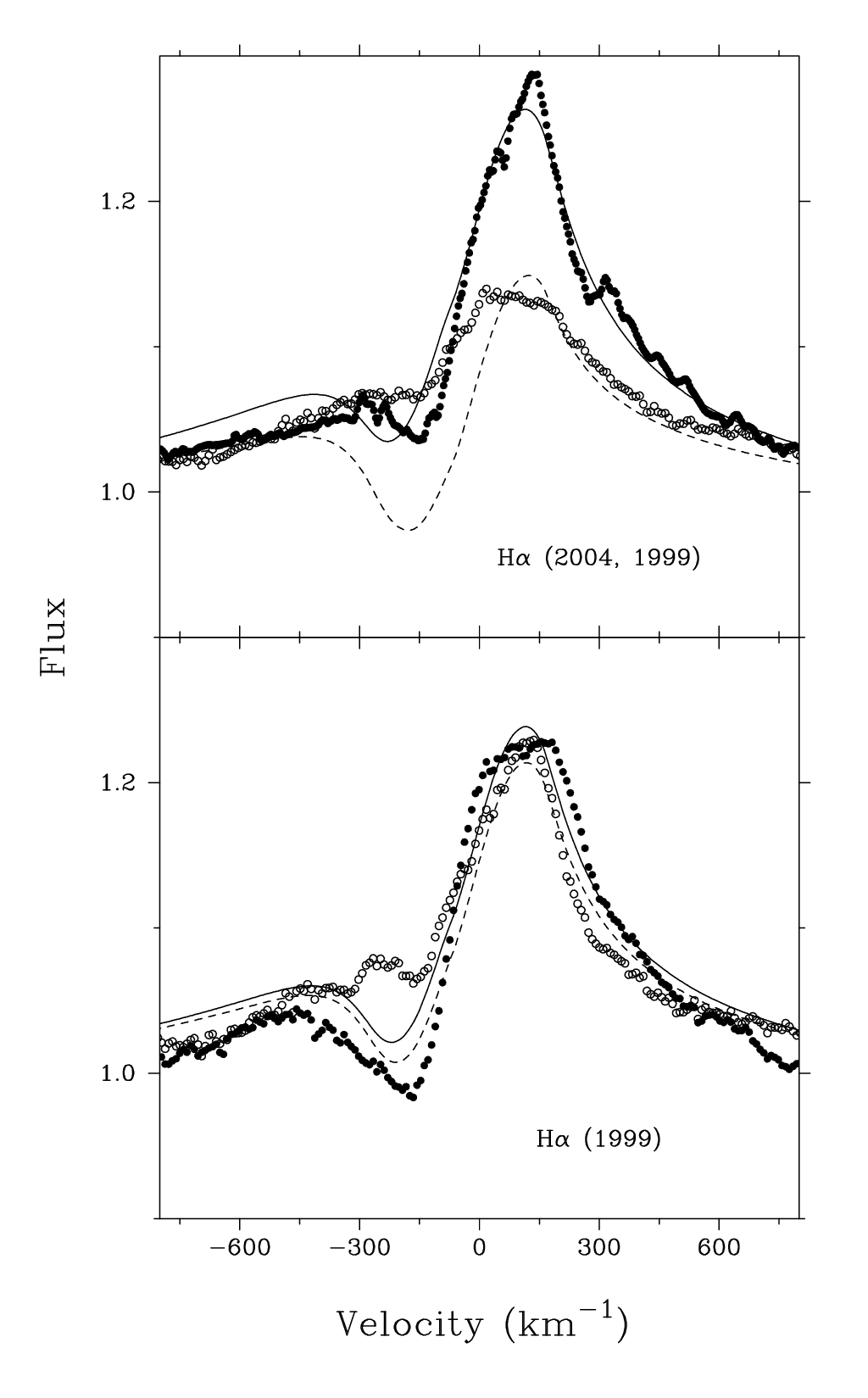}
   \caption{\Ha model fits (solid and dashed lines) to extreme 
changes observed in 2004 and 1999 (filled and open circles).}
              \label{fig8}%
    \end{figure}
%

Hot massive stars are expected to be subject to photospheric instabilities
and/or pulsations.  During the last decade the  search for non-radial pulsations 
in OB stars has been largely motivated by the need to test the hypothesis of the 
``photospheric connection" which aims at linking  the cyclic variability 
observed in the winds of these  stars to processes in the stellar 
photosphere. The task however turned out to be  observationally very 
demanding  and as a result  only a limited number of OB stars are 
currently flagged as confirmed or suspected pulsators 
\citep[e.g.][]{Fullerton96, 
Reid1996, deJong01, Kaufer06}.

The results derived in this study {raise the possibility} that the 
supergiant $\alpha$~Cam 
might also be a non-radial pulsator.  Radial velocity variations  in  optical 
absorption lines of $\alpha$~Cam have been reported previously by  
\citet{Zeinalov86} and \citet{ Musaev88}  who argued
that these variations are 
likely  periodic  with a possible cycle length of 1.35, 1.54 or 2.93 days 
(with no preference to any of them).  However, our analysis of the  
2002 and 2004 time-series  did not support any of  these periods but 
instead revealed a new one of 0.36 days, which may be interpreted in terms of 
prograde low-order non-radial pulsational behaviour.
{The $\sim$ 8--9 hour modulation may persist over two months in 2002,
albeit with different phase shifts.} This signal is not evident in
data collected more than 2 years later 
(November 2004); it is also not seen in the stellar wind components of the line 
profiles. {The interpretation of the 0.36-day  variation in terms  
of  photospheric 
velocity fields is also  supported  by weak fluctuations in the
shallow photospheric lines of C~IV $\lambda\lambda$~5801, 5812.}

Apart from the 0.36-day periodic variation detected between  Jan- March 
2002, our observations also documented occasional  large-amplitude 
(of about 15 to 30 \kms) differential radial velocity shifts in \he and \hee
absorption lines, accompanied by an increase in the \Ha emission flux 
centred at \vsys and localised within $\pm$ \vesini. 

Large-amplitude differential  velocity shifts have been discovered 
by \citet{Prinja04} in the optical absorption lines of $\epsilon$ Ori (B0 Ia) 
and they are also present  in the photospheric lines of $\beta$ Cepheid 
stars. However in both these cases the velocity shifts are regular and periodic 
while those seen in  $\alpha$~Cam occurred over at least several  hours, during
a specific night in 2004,  with no recurrence during  the other nights of the run.
This phenomenon can be interpreted either in terms of large-amplitude changes 
in photospheric velocity fields or as the result of deep-seated, 
large-scale wind structure that 
occasionally modulates the density distribution in the innermost
part of the wind.

Discrete Absorption Components moving from red to blue across  
the unsaturated  UV resonance lines are the
characteristic signature of large 
scale structure in O star winds. Unfortunately the UV resonance lines 
of $\alpha$ Cam  are saturated and do not provide any detailed information 
on variable wind features \citep{Kaper1997}. We have re-examined 
the $IUE$ time-series of $\alpha$ Cam from February 1991 and December 1994.
Whilst there is subtle evidence for low velocity fluctuations in
Si~IV $\lambda\lambda$1400 (and Si~III $\lambda$1207), the clearest
variability is in the soft blue wings of the saturated UV lines.
For example, during the 1991 $IUE$ observing run of $\sim$ 5 days, 
additional absorption is evident in the blue wing of Si~IV beyond $v_\infty$
(= 1550 km s$^{-1}$) out to $\sim$ $-$1700 km s$^{-1}$. This enhancement
persists for $\sim$ 24 hours and does not recur during the following
4 days. This finding is at least consistent with the notion based earlier  
on optical observations, that large scale structure may develop in 
the wind of $\alpha$ Cam; the present study suggests however that
these structures are somewhat unstable and thus do not give rise 
to longer-term periodic phenomena in the wind.

The possibility that the wind of $\alpha$ Cam is not smooth but
structured has got additional support by the recent results derived by 
\citet{Fullerton06} and \citet{Puls06}. In particular,  \citet{Fullerton06} have 
modelled the P~V lines of $\alpha$ Cam using the SEI method to derive 
products of mass-loss rate and ionization fraction. Adopting mass-loss rates 
derived from radio observations, they find resulting ionization fractions, 
q(P$^{4+}$), = 7$\times$10$^{-3}$ for $\alpha$ Cam. In the case of 
$\alpha$ Cam the low ion fraction is highly discrepant with results from NLTE 
stellar atmosphere models which predict that the P~V ion fraction should 
be 
dominant and thus close to unity at these effective temperatures. 
The discrepancy is thought to indicate that the wind is highly clumped 
and has very small volume filling factors. 

Further evidence in support of the clumped structure of $\alpha$~Cam wind
has been derived by \citet{Puls06}. By means of a  simultaneous model  
analysis of \Ha, IR and radio data these authors  showed  that the wind of 
$\alpha$ Cam likely has a clumping factor in the innermost region ( between 
1.05  \footnote{The wind below 1.05\Rstar was assumed to be unclumped 
in agreement with results from hydrodynamical simulations which showed that 
the line-driven  instability needs some time to grow before significant structure is 
formed.} and  2\Rstar),  being a factor of 2.6 larger than in the outermost wind 
(beyond 50\Rstar). Using the \Ha emission wings as a tracer of wind clumping in 
the intermediate wind (between 2 and  15\Rstar), Puls et al. furthermore found 
that the clumping factor in this region is similar or fractionally larger than in 
the inner wind. 

Interestingly, the properties of the TVS of \Ha (derived in the present 
study  and by \citet{Markova02}) also seem to be consistent with results 
obtained via line-profile simulations including wind clumps \citep{Markova05}. 
In particular, the velocity limits as well as the established asymmetry of 
the TVS are both consistent with a model  including 'broken shells" 
(i.e.~ clumps), uniformly  distributed over the wind volume, with density 
contrast  $\delta \rho/\rho_0$ of $\pm$0.35 (model series BS21 in
\citep{Markova05}) where $\rho_0$ is the stationary density.  
Thus, it may be  that \Ha night-to-night variability observed in 
$\alpha$ Cam 
is at least partially connected with presence of small-scale structure (clumps, 
blobs) in the innermost part of the wind.

Overall, there is now substantial observational evidence which suggests 
that the wind of $\alpha$ Cam is not smooth but structured over, both, 
large and 
small spatial scales.  In view of this possibility it is therefore not surprising
 that we failed to reproduce the observed set of \Ha profiles (Sec. 4) by means 
of model calculations. Indeed, the approximate method of \citet{Puls1996} employed 
here, as well as most of the state-of-art model atmosphere codes,  rely on the 
assumption of a globally stationary wind with a smooth density /velocity 
stratification, {coupled to a `stable' photospheric line contribution}, 
and are therefore incapable in principle of describing 
the observed {\it blue-ward (absorptive)} behaviour of \Ha in $\alpha$ Cam 
and of any other star with a substantially structured wind.

\acknowledgement{
This study was in part supported by funds from the National Scientific 
Foundation to the
Bulgarian Ministry of Education and Science (F-1407/2004). We are also grateful to 
Bulgarian Academy of Sciences and the Royal Society (UK) for a collaborative 
research grant.  NM acknowledges the financial support of 
the Bulgarian 
Academy of Sciences and PPARC. NM is also grateful to her colleague 
Sabotinov (BAS) for his 
understanding and moral support during this study.}

\end{document}